\makeatletter
\def\input@path{}
\makeatother

\documentclass[twocolumn,epjc3]{svjour3}  

\RequirePackage{StrInJetUE}
\smartqed  
\journalname{Eur. Phys. J. A}
\begin{document}
\begin{sloppypar}
\title{Studying strange hadron productions via two-particle correlations in high energy proton-proton collisions}


\author{M.N. Anaam\thanksref{e1, addr1}
        \and
        Liang Zheng\thanksref{e2, addr2, addr1}
        \and 
        H.M. Alfanda\thanksref{e3, addr1}
        \and
        Zhongbao Yin\thanksref{e4, addr1}
}

\thankstext{e1}{e-mail:mustafa.naji.anaam@cern.ch (corresponding author)}
\thankstext{e2}{e-mail:zhengliang@cug.edu.cn}
\thankstext{e3}{e-mail:haidar.mas'ud.alfanda@cern.ch}
\thankstext{e4}{e-mail:zbyin@mail.ccnu.edu.cn}


\institute{
Key Laboratory of Quark and Lepton Physics (MOE) and Institute of Particle Physics,
Central China Normal University,
Wuhan 430079, China \label{addr1}
   \and
School of Mathematics and Physics, China University of Geosciences (Wuhan), Wuhan 430074, China \label{addr2}
}

\date{Received: date / Accepted: date}

\maketitle
\begin{abstract}
Strange hadron production in \pp collisions at \seven is studied with \Pyeight and \Epos event generators via the two-particle correlation method. After pairing charged particles as trigger particles with associated strange hadrons (\kzero and \lmb(\almb)), the strange baryon to meson per trigger yield ratios ${Y_{\Delta\varphi}^{\mathrm{h}-(\lmb+\almb)}}
/{2Y_{\Delta\varphi}^{\mathrm{h}-\mathrm{K}^{0}_{S}}}$ dependent on transverse momentum (\pt) are investigated within the near- and away-side jet region as well as the underlying event (UE) region. The modified string fragmentation effects of color reconnection and string shoving mechanisms implemented in \Pyeight and the final state evolution effects built in \Epos are compared in this study. It is found that \pt dependence of the \lmb/\kzero per trigger yield ratio in UE is similar to that in the inclusive measurements while the in-jet result is smaller. We also show that the \lmb/\kzero ratio at both near- and away-side jet region in the intermediate \pt region decreases significantly with trigger \pt, which suggests the hard process effects might contribute to the strange baryon to meson enhancement at intermediate \pt. The much higher \lmb/\kzero ratio in UE and the striking difference between the near and away side of the in-jet results with low trigger \pt  predicted by the \Epos model indicate that the soft process effects induced by the final state evolution further increase the strange baryon production and can be still effective in a higher \pt region compared to the string fragmentation type models.
\end{abstract}
\section{Introduction}
\label{sec:intro}
The transverse momentum \pt dependence of the strange baryon to meson \lmb/\kzero ratio provides a unique way to study the properties of the hot and dense, color-deconfined QCD matter called quark-gluon plasma (QGP) created at the Relativistic Heavy Ion collider (RHIC) and the Large Hadron Collider (LHC). It is observed that a pronouncing enhancement of the strange baryon to meson ratio at intermediate \pt exists in nucleus-nucleus (\nucnuc) collisions compared to proton-proton (\pp) collisions~\cite{ALICE:2013cdo}. It is believed that this anomaly strange baryon enhancement and the shift of the \lmb/\kzero peaks towards higher \pt can be described by the introduction of the parton recombination mechanism to the hadronization process and the hydrodynamic parton radial flow induced due to the formation of the QGP phase. However, the enhanced strange baryon to meson ratio at intermediate \pt, together with many other QGP like phenomena~\cite{ALICE:2016fzo,CMS:2010ifv,CMS:2016fnw,ALICE:2018pal}, has also been observed in high multiplicity \pp and \pPb collisions~\cite{ALICE:2020wfu,ALICE:2016fzo,ALICE:2021npz,ALICE:2013wgn}.

The similarity between the small systems and the large systems invites a lot of discussions on whether the QGP matter is formed in high multiplicity \pp and \pPb collisions at the LHC. Different explanations are modeled in the general purpose Monte Carlo event generators like \Pyeight~\cite{Sjostrand:2014zea} and \Epos~\cite{Pierog:2013ria}. In \Pyeight, modified string fragmentation models considering the interactions between different strings created in the multiple parton interactions are implemented via the color reconnection (CR) and string shoving mechanism~\cite{Bierlich:2015rha,Bierlich:2016vgw}. The \Epos model introduces the final state hydrodynamic evolution to the small system collisions by assuming that the high density region is thermalized within the core-corona picture. These models can offer similar strange baryon enhancement in the inclusive \lmb/\kzero ratio at intermediate \pt in high multiplicity \pp or \pPb collisions. To further pin down the origin of this behavior in small systems, it is important to separate the contributions from soft and hard processes. The strange hadron productions have been studied with reconstructed high-\pt charged particle jets to investigate the contribution of fragmentation of fast moving partons~\cite{ALICE:2021vxl}. The enhancement of charm baryon to meson ratio (\lmbc/\Dzero) is also observed in high multiplicity \pp collisions ~\cite{ALICE:2021npz}. Considering the charm hadrons are mostly produced through hard processes, the resemblance of \lmbc/\Dzero to the strange sector \lmb/\kzero uncovered in the experiment indicates the existence of modifications to the hard parton hadronization process in small systems.  

Studying strange hadron correlations with other charged hadrons in different kinematic ranges can be useful to expose the effects of jet fragmentation, parton shower and hydrodynamic evolution~\cite{ALICE:2021nvv}. From previous studies, it is known that high-\pttrig particles are more likely to come from hard processes, such as jet fragmentation, while at intermediate \pttrig region the particle yield contains also contributions from soft processes like multiple parton interactions, parton radiation effects which cannot be easily separated in experiments~\cite{Strangeness1,Strangeness2,ALICE:2016fzo,ALICE:2021npz,ALICE:2020wfu}. Particles from jet fragmentation are more collimated with the jet axis, resulting in the near-side $(\dphi=0)$ and the back-to-back away-side $(\dphi=\pi)$ peaks, where minijet contributions with the momentum conservation are dominant~\cite{1993}. Particles from the soft production process associated with the underlying event (UE) region of the correlation function are more isotropic and only have some minor sensitivity to the hard process. One may expect significant differences between the strange hadron productions associated with low \pttrig and high \pttrig particles due to the large variations of the $z$ (fraction of parton momentum) regions probed. 

The goal of this study is to analyze the strange baryon and meson production in the peak and UE regions in \pp collisions in order to map out the underlying physics mechanisms. The relative contribution of hard and soft processes to strangeness production in \pp collisions is still not clear and can be studied through techniques involving the two-particle correlations. We explore the baryon-to-meson anomaly at low and high \pttrig to determine whether it is arising from the soft, collective, UE type physics, or the hard processes involving the jet fragmentation. The study is performed on \kzero, \lmb/\almb with transverse momentum within \ptassrange{1}{6} associated with charged hadrons as trigger particles having transverse momentum within \pttrigrange{1}{6} and \pttrigrange{6}{15}.

This paper is organized as follows: section~\ref{sec:generators} provides details on the event generators, section~\ref{sec:correl} provides definition of the observable, section~\ref{sec:results} is devoted to results and discussion. Finally, the paper is summarized in section~\ref{sec:summary}.\\

\section{PYTHIA and EPOS LHC event generators}\label{sec:generators}

\Py is a general purpose perturbative-QCD based Monte Carlo event generator, widely used for high-energy collider physics with an emphasis on \pp collisions. \Py is quite successful in describing various experimental results of \pp collisions at LHC energies~\cite{ALICE:2013rdo,ALICE:2015olq,ALICE:2016jjg,ALICE:2015olq}. For a better description of data, different parameters of the existing \Py model have been fine-tuned over time.

In this work, the study is performed using \Pyeight (version 8.2.4) with hadronization models considering CR and shoving. The hadronization in \Py is treated using Lund string fragmentation model~\cite{Metz:2016swz,Christiansen:2015yqa}. In the CR picture of the model~\cite{Argyropoulos:2014zoa,Bierlich:2015rha}, the final partons are considered to be color connected in a way that the total string length becomes as short as possible~\cite{Gustafson:2009qz}. Within the multiple parton interaction (MPI) framework~\cite{Fedkevych:2020cmd,Diehl:2017wew} implemented in \Pyeight, a large number of final state string objects can be generated over limited transverse space in high energy collisions. The traditional string fragmentation process is expected to be modified in the high parton density environment due to the inter-string interactions~\cite{Christiansen:2015yqa,Sjostrand:2014zea}.

The parameters used in this study are the same of new parameters developed for LHC in Ref~\cite{Bierlich:2015rha} with respect to the default settings of the Monash Tune.

String shoving model is recently proposed to describe the strangeness enhancement and long-range correlation in high multiplicity \pp collisions. The model is implemented in the DIPSY event generator. It introduces a repulsive force which allows the strings overlapped in transverse space to act coherently and generate a flow-like effect in small systems~\cite{Bierlich:2016vgw}.

\Epos is a Monte Carlo event generator for minimum bias hadronic interactions, used for both heavy-ion interactions and cosmic ray air shower simulation. The \Epos employs the core-corona picture which describes high energy collisions ranging from small systems to large systems in a unified manner~\cite{Pierog:2013ria}. The high density areas are referred to as the core, the low density areas as the corona~\cite{Werner:2007bf}. The inclusion of the thermalized core area of the produced medium brings a parameterized hydrodynamic evolution effect to the system. More details about \Epos used in this work can find in the Ref.~\cite{Pierog:2013ria}. 

\section{Correlation function and acceptance correction}
\label{sec:correl}

The correlation function is defined as a function of the azimuthal angle difference $\dphi=\phitrig-\phiass$ and pseudorapidity difference $\deta=\etatrig-\etass$ between trigger and associated particles. Pair acceptance corrections are evaluated with a mixed-event technique, where (\dphi,\deta) distributions with the trigger and the associated particles originated from different events are constructed. The correlation function $C(\dphi,\deta)$ is extracted using the following procedure to correct for acceptance effects:

\begin{equation}
  C(\dphi,\deta)=B(0,0)\frac{S(\dphi,\deta)}{B(\dphi,\deta)}\quad,
    \label{correlation 2D}
\end{equation}
    
  \noindent $S(\dphi,\deta)$ is constructed from particle pairs coming from the same event.

\begin{equation}
   S(\dphi,\deta)=\frac{1}{\Nsig}\frac{\dd^{2}\Nsig}{\dd\dphi\dd\deta}\quad,
   \label{correlationsame 2D}
\end{equation}

\noindent where \Nsig is the number of pairs of particles in $S$. $B(\dphi,\deta)$ is constructed using an event mixing technique, where each particle in the pair comes from a different event and can be expressed as

\begin{equation}
   B(\dphi,\deta)=\frac{1}{\Nmix}\frac{\dd^{2}\Nmix}{\dd\dphi\dd\deta}\quad,
   \label{correlationmix 2D}
\end{equation}
  
\noindent where \Nmix is the number of pairs of particles in $B$. The $B(0,0)$ is the per-trigger yield at $(\Delta\varphi,\Delta\eta)=(0,0)$, where the pair acceptance is maximum~\cite{K0sandLambda_PbPb,Multiplicitydependence,Long_range_correlations_PPb}.

\begin{figure}[ht!]%
\centering
\includegraphics[width=8cm]{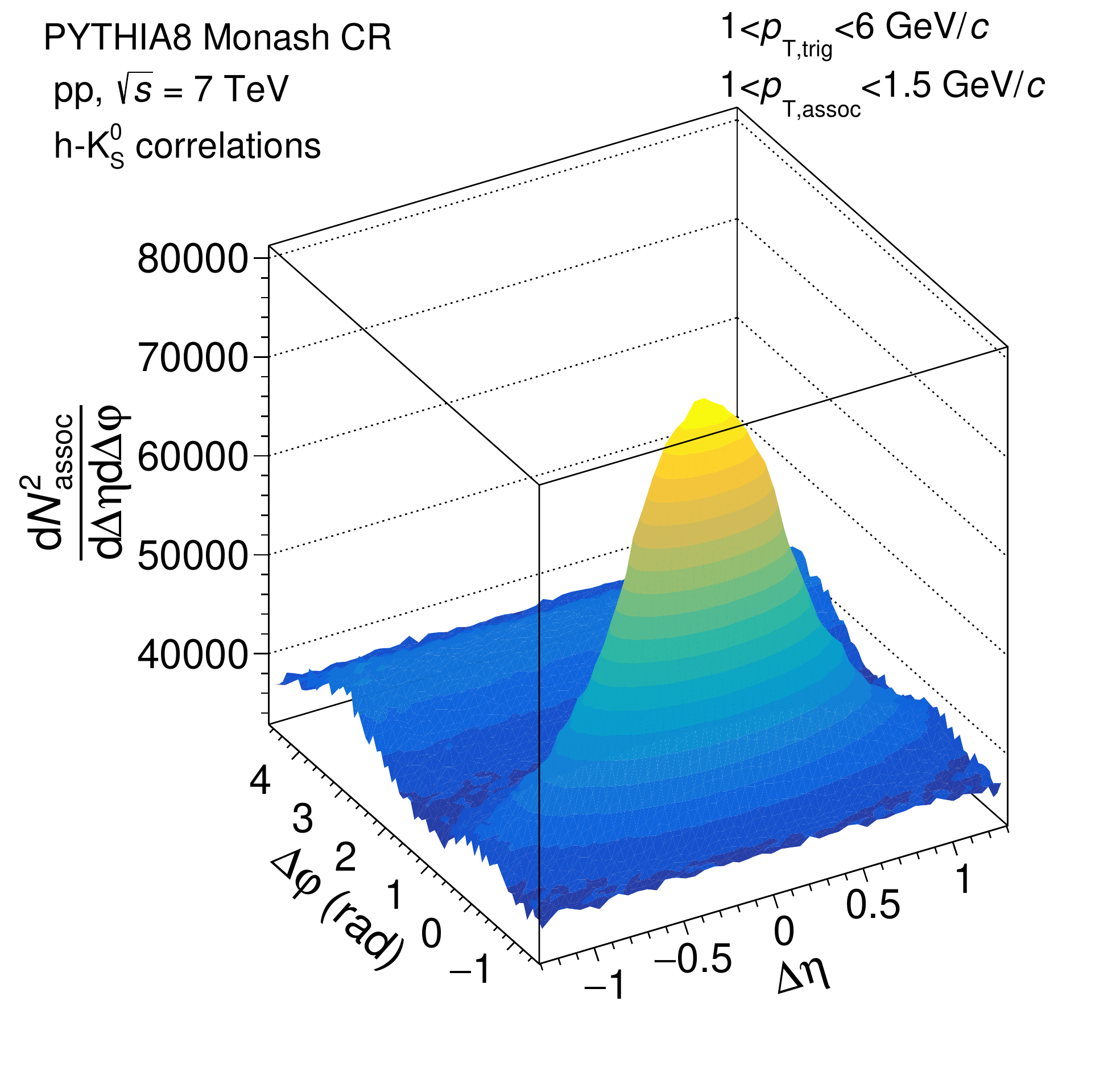}
\includegraphics[width=8cm]{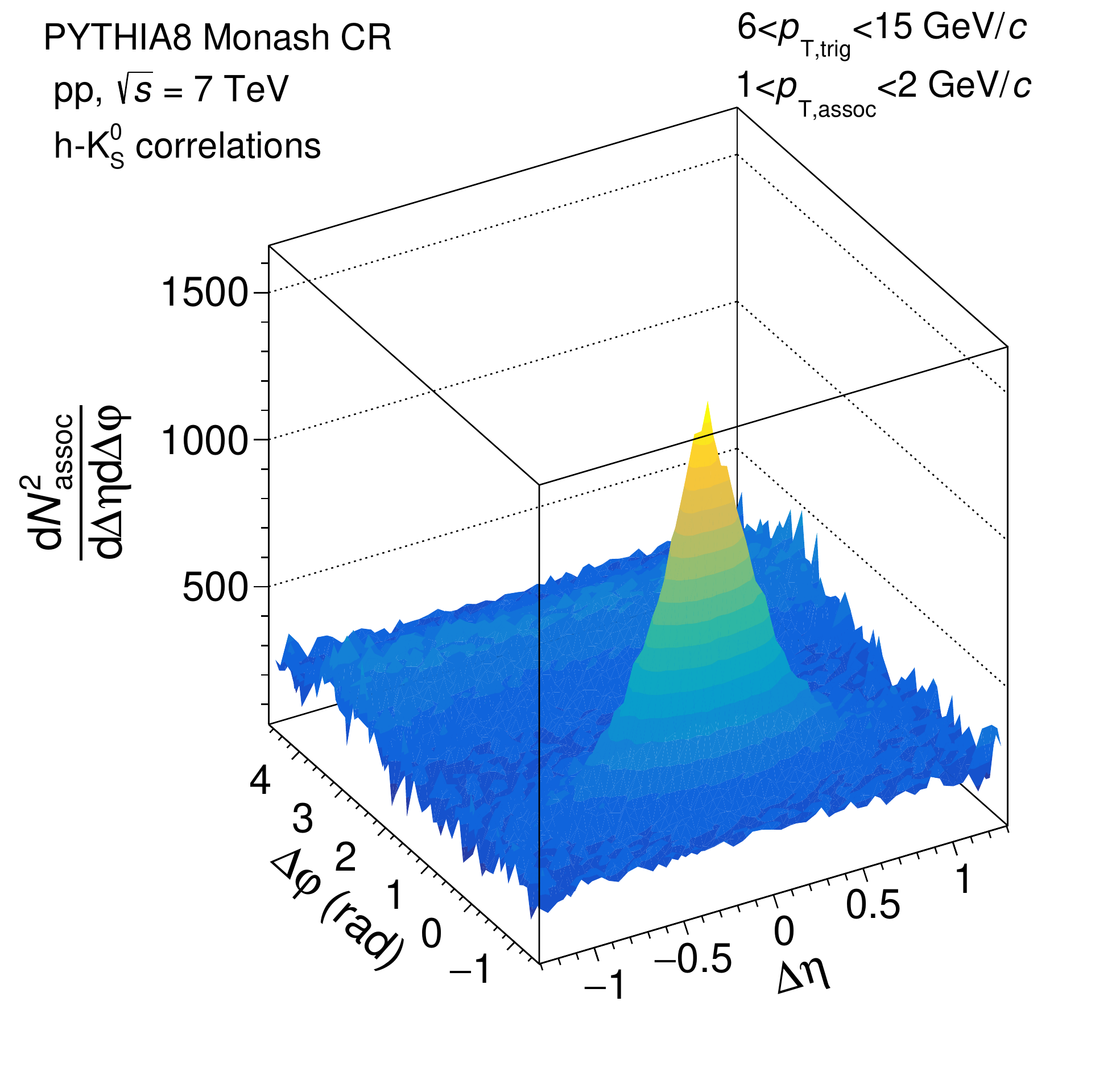}
\caption{h--\kzero correlation functions after acceptance correction in \pp collisions at \seven with \pttrigrange{1}{6} and \ptassrange{1}{1.5} (upper panel), and \pttrigrange{6}{15} and \ptassrange{1}{2} (bottom panel) using \Pyeight CR.}
\label{fig.acceptance2d1}
\end{figure}

$C(\dphi,\deta)$ with \Pyeight CR is shown in Fig.~\ref{fig.acceptance2d1} for h--\kzero correlation in \pp collisions of \ptassrange{1}{1.5} and \ptassrange{1}{2} associated with \pttrig transverse momentum \pttrigrange{1}{6}, and \pttrigrange{6}{15}, respectively. The near-side peak of two particle correlation in \pp collisions in  $(\Delta\varphi,\Delta\eta)$ about $(0,0)$ is observed similar to the results at LHC energies~\cite{CMS:2010ifv,CMS:2015fgy,ATLAS:2015hzw}, which is a combination of at least three effects, fragmentation of hard-scattered partons, resonance decays, and femtoscopic correlations. The fragmentation originating from low momentum-transfer scatterings, sometimes referred to as minijets~\cite{ALICE:2014mas}, produces a broad structure extending at least over one unit in $\deta$ and $\dphi$.

\section{Results and discussion}%
\label{sec:results}

The result of $\dphi$ distribution, the per-trigger yields of \kzero and (\lmb+\almb), and their ratios in \pp collision at \seven within \ptassrange{1}{6} associated with primary charged particles of \pttrig intervals within \pttrigrange{1}{6} and \pttrigrange{6}{15} will be discussed in this section.

\newpage
\subsection{Projection of correlation function }

The per-trigger yield of associated strange hadrons is studied as a function of the azimuthal angle difference $\dphi$ in \pp collisions. This distribution is given by

\begin{equation}
    C(\Delta\varphi)=\frac{1}{\Ntrig}\frac{\dd \Nass}{\dd\Delta\varphi}\quad.
\end{equation}

\begin{figure*}[ht]
	\begin{center}
	\includegraphics[width=0.95\textwidth]{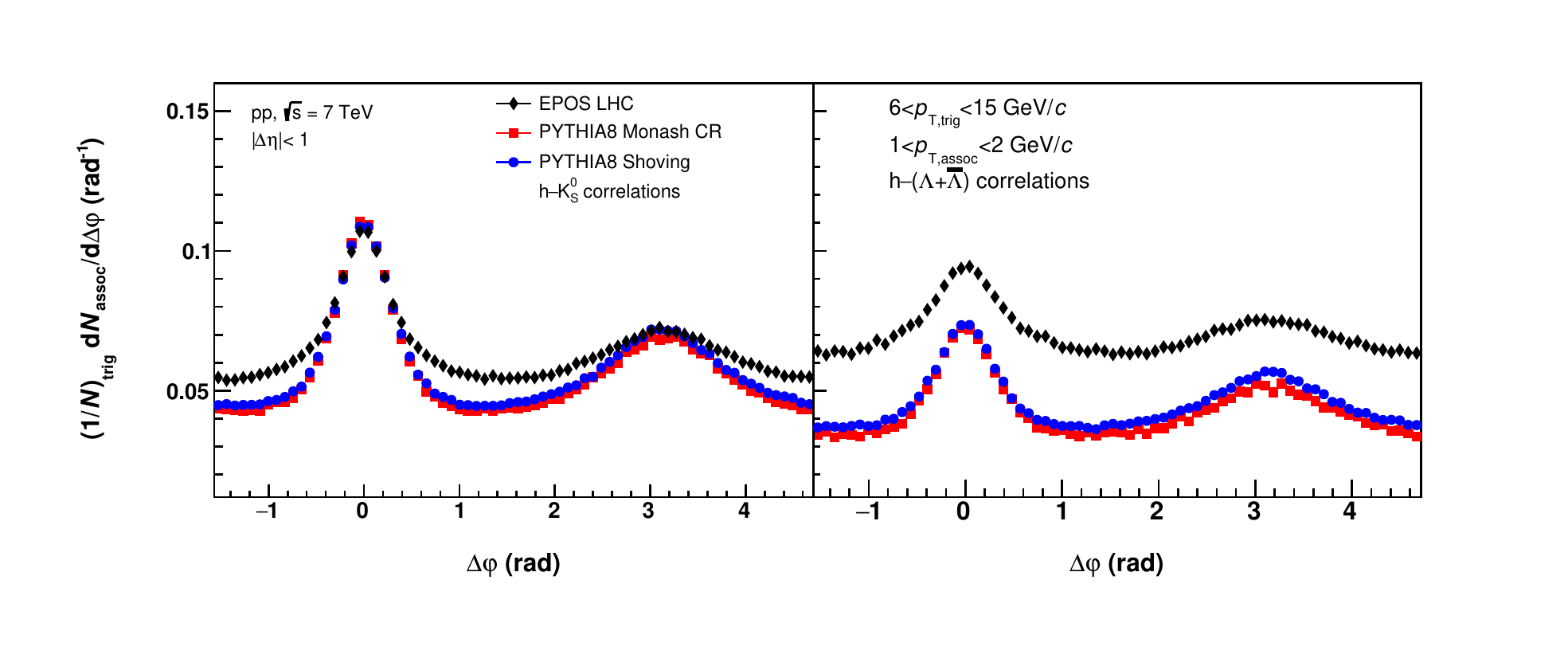}
	\includegraphics[width=0.95\textwidth]{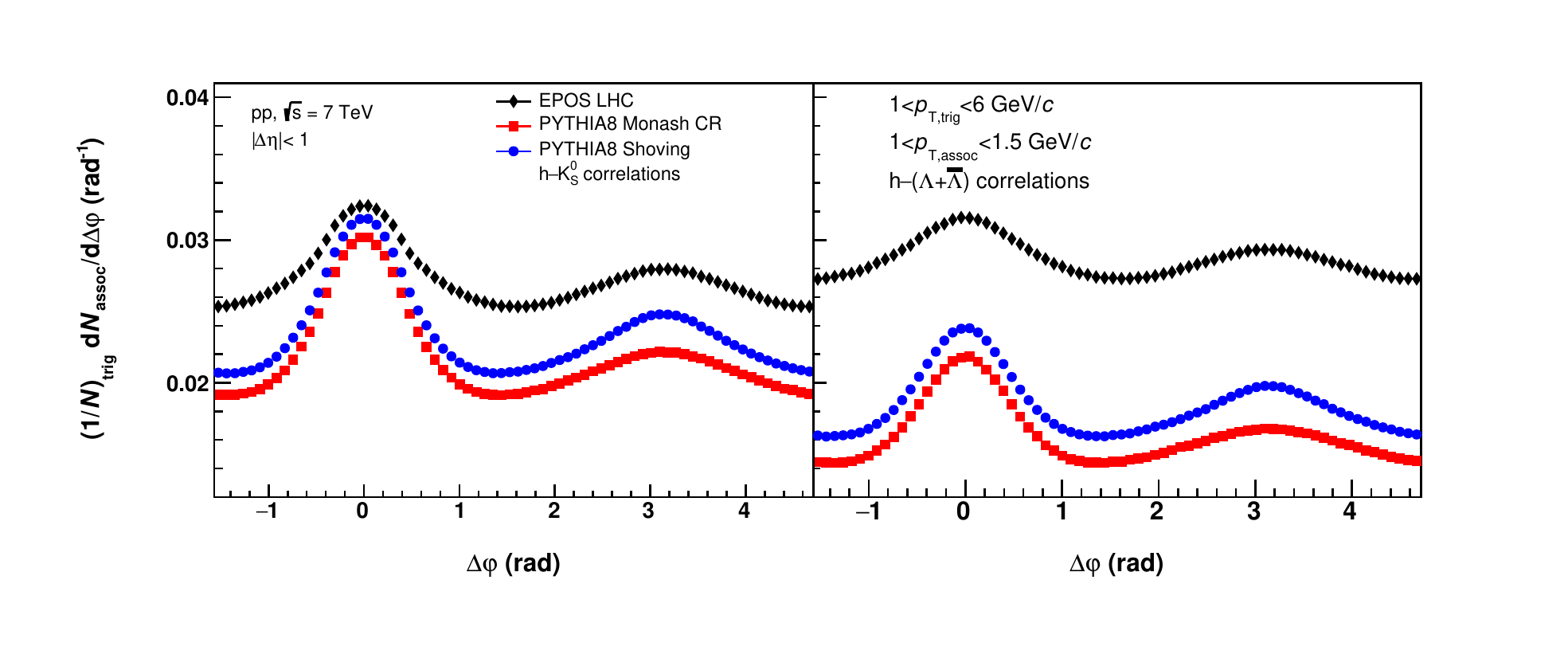}
	\end{center}
    \caption{Correlation function distributions $C(\Delta\varphi)$ for h--\kzero (left column) and h--($\lmb+\almb$) (right column) with \pttrig range \pttrigrange{6}{15} (top raw) and \pttrigrange{1}{6} (bottom raw) and corresponding \ptass range \ptassrange{1}{2} and \ptassrange{1}{1.5}, respectively. Different marker style correspond to different model (\Epos is shown in diamond, PYTHIA8 Monash CR in square, and \Pyeight shoving in circle.}
	\label{fig:dphi}
\end{figure*}

\noindent This quantity is presented for pairs of particles where $\ptass<\pttrig$ within $\abs{\deta}< 1$ as a function of \ptass. $C(\Delta\varphi)$ distributions of  \kzero and $(\lmb+\almb)$ 
particles for different \ptass intervals with \Pyeight CR, \Pyeight shoving and \Epos are shown in Fig.~\ref{fig:dphi} in \pp collisions at \seven before subtracting the underlying event contribution by the zero yield at minimum (ZYAM) method~\cite{ZYAM_flow}.

\subsection{Per-trigger yields }

The per-trigger yields on the near side and away side are obtained by subtracting the underlying event contribution and integrating $C(\Delta\varphi)$ in the ranges of $\abs{\dphi}<1$ and $\abs{\dphi-\pi}<1.2$, respectively. The UE yield is also obtained by integrating $C(\Delta\varphi)$in the interval of $1.3 <{\dphi}<1.8$.

\begin{figure*}[ht!]
	\begin{center}
	\includegraphics[width=1\textwidth]{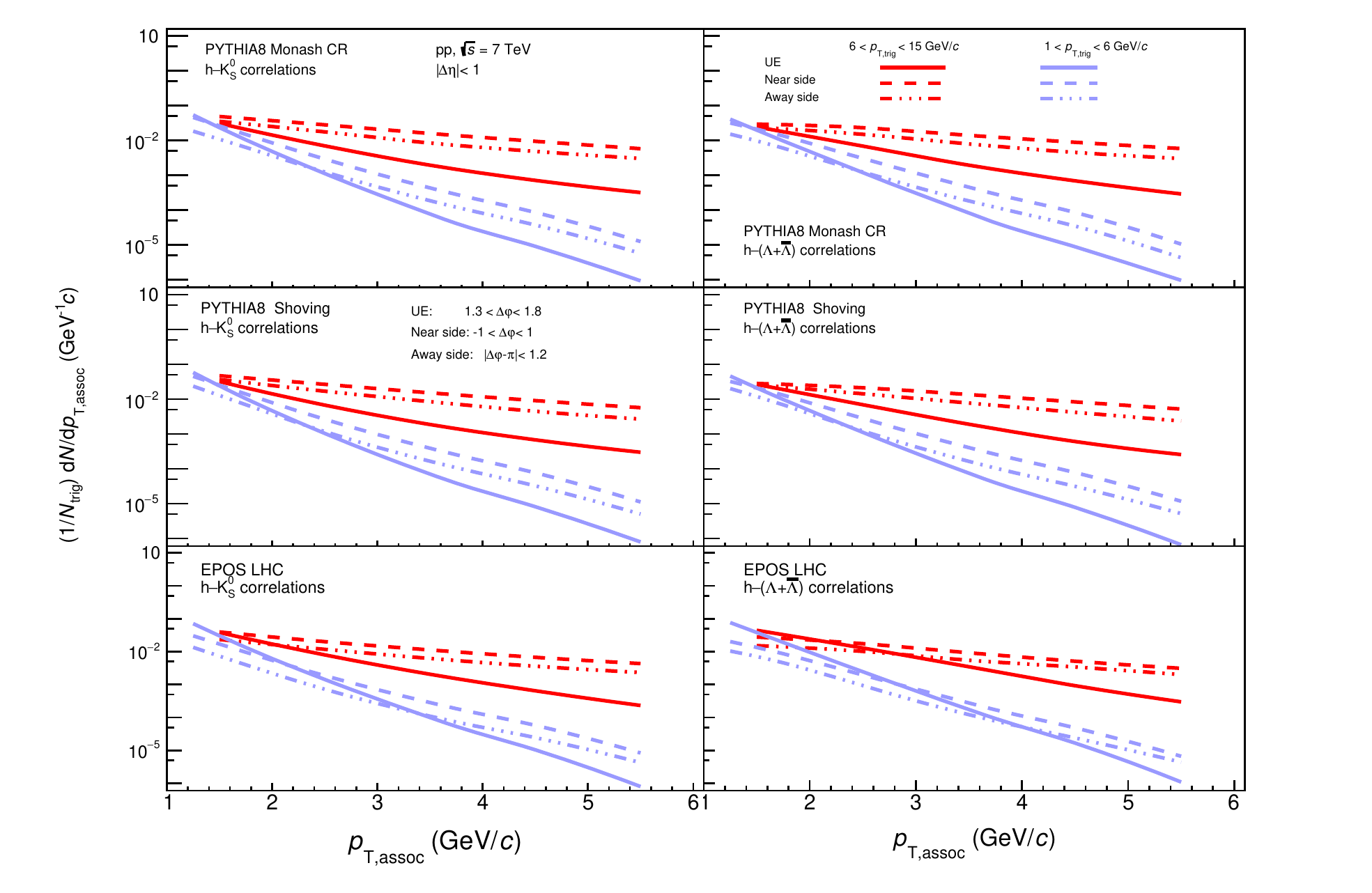}
	\end{center}
	\caption{Per-trigger yields in the UE (full solid lines), near-side jet (dashed lines) and away-side jet (dashed-point lines) in the \pttrig range \pttrigrange{6}{15} (red marker) and \pttrigrange{1}{6} (blue marker) of \kzero on the left and $(\lmb+\almb)$ on the right, with \Pyeight CR (top row), \Pyeight Shoving (middle row) and \Epos (bottom row).}
	\label{fig:per_trigg_yield}
\end{figure*}

The per-trigger yields for the near- and away-side peaks and UE region as a function of the \ptass of \kzero and $(\lmb+\almb)$ are presented in Fig.~\ref{fig:per_trigg_yield} in \pp collisions at \seven for two \pttrig intervals within \pttrigrange{1}{6} and \pttrigrange{6}{15}.
The results show that the yield decreases with increasing \ptass. It also show that the difference between the two \pttrig yields becomes large with increasing \ptass suggested that the higher energetic jets have increasingly more associated particles.

The per-trigger yields associated with \pttrigrange{6}{15} are always harder in jet than that in UE, specially for $\ptass>2.5$~\GeVc. On the other hand, the per-trigger yields associated with \pttrigrange{1}{6} become more prominent in UE for $\ptass>2.5$~\GeVc.

The per-trigger near-side and away-side yields provide information about fragmentation properties of low-\pt partons.

\subsection{ Baryon to meson per trigger yield ratio}

\begin{figure*}[ht!]
	\begin{center}
		\includegraphics[width=1\textwidth]{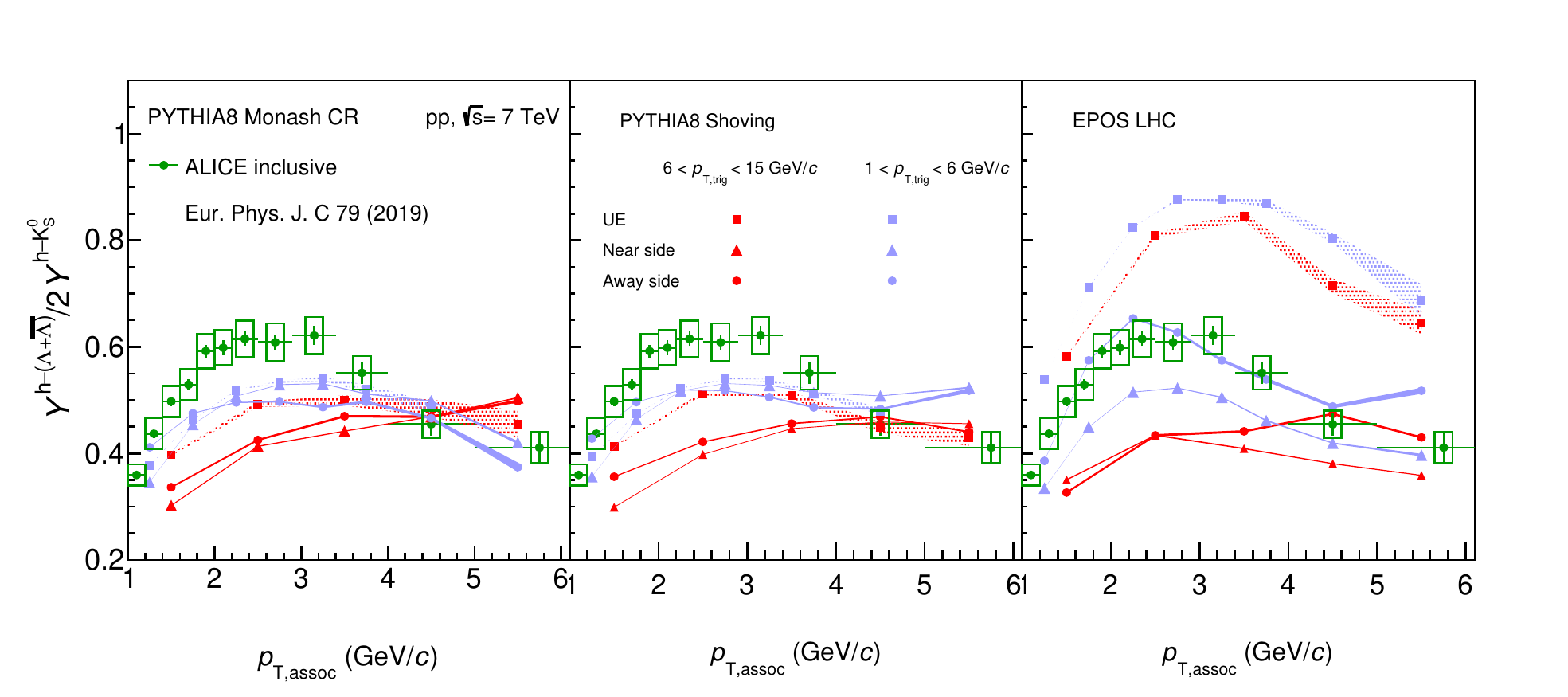}
	\end{center}
	\caption{Per-trigger yield ratios, ${Y_{\Delta\varphi}^{\mathrm{h}-(\lmb+\almb)}}
/{2Y_{\Delta\varphi}^{\mathrm{h}-\mathrm{K}^{0}_{S}}}$, in the UE, near-and away-side jets of \pttrigrange{6}{15} and \pttrigrange{1}{6} in \pp collisions at \seven with \Pyeight CR (left panel), \Pyeight shoving (middle panel), and \Epos (right panel).}
\label{fig:ratios}
\end{figure*}

The baryon to meson per trigger yield ratio ${Y_{\Delta\varphi}^{\rm{h}-(\lmb+\almb)}}
/{2Y_{\Delta\varphi}^{\rm{h}-\kzero}}$ is investigated as a function of transverse momentum of the associate strange particles in order to compare the production of strange hadrons in jets and UE. Figure~\ref{fig:ratios} shows this ratio in two \pttrig intervals \pttrigrange{1}{6} and \pttrigrange{6}{15} studied with the \Pyeight CR, Shoving and \Epos event generators. The ALICE inclusive strange baryon to meson ratio~\cite{ALICE:2020jsh} varying with \pt has also been displayed in the figure as a guidance. Similar to the \lmb/\kzero ratio shown in the inclusive data, the per trigger yield ratios in jet and UE from the correlation method are initially increasing with \ptass and reach a plateau at around $2.5$~\GeVc.
The UE per trigger yield ratios in \Pyeight model are close to the inclusive data results, while the \Epos model over-predicts strongly \lmb/\kzero ratio in the UE region. It is indicated by this comparison that the deconfined quark matter evolution effects embedded in \Epos lead to a much stronger strange baryon enhancement compared to the multi-string interaction effects in \Pyeight. On the other hand, the in-jet results with \pttrigrange{6}{15} are found to be quite similar across all three different model calculations. This similarity suggests that the high \pttrig in-jet results are completely determined by the jet fragmentation effects and independent of the soft physics implementations in the models. More interestingly, it is noted that the in-jet \lmb/\kzero results with \pttrigrange{6}{15} are lower than those with \pttrigrange{1}{6} in the \ptass region less than $4$~\GeVc for all model calculations. This \pttrig dependence uncovers the importance of hard process effects from minjet productions to the strange baryon to meson enhancement at intermediate \pt region. 
Study of strange hadron productions associated with low- and high-\pt charged particles will help us to disentangle the strange particles produced in hard and soft processes.

\section{Summary}%
\label{sec:summary}

Two-particle azimuthal correlations of h--\kzero and h--$(\lmb+\almb)$ are studied in \pp collisions at \seven to explore the jet fragmentation effects and the interplay between soft and hard process with \Pyeight (CR and Shoving) and \Epos event generators. The ratios of the per-trigger yields ${Y_{\Delta\varphi}^{\mathrm{h}-(\lmb+\almb)}}/{2Y_{\Delta\varphi}^{\mathrm{h}-\mathrm{K}^{0}_{S}}}$ are calculated within jets and UE region to understand the relative contribution of different mechanisms to the strange baryon to meson enhancement at intermediate \pt. It is found that strange baryon to meson ratio of the per-trigger yield in the UE region is similar to the measured inclusive results, while the in-jet ratios are generally smaller than the UE results. However, the in-jet ratio is enhanced from high \pttrig to low \pttrig in the region of \ptass less than $4$~\GeVc, and it becomes similar in the high \pttrig bin across all three different model calculations. This behavior indicates that the hard process jet fragmentation effect is important for the understanding of the strange baryon to meson ratio in the intermediate \pt region. For high enough \pt such as above $4$~\GeVc, the hard process effect is dominant. 

With the string fragmentation based models implemented in \Pyeight, the in-jet per trigger yield ratios are mildly suppressed within the \ptass range from  $1$~\GeVc to $4$~\GeVc in the high \pttrig bin compared to the low \pttrig bin and no significant difference can be found at the near side and the away side. The \Epos model predicts a much higher \lmb/\kzero ratio in UE compared to the string fragmentation models. The in-jet ratio is found to be larger at away side in the \pttrigrange{6}{15} bin. This behavior suggests the soft process effects modeled through the final state evolution effects strongly increase the strange baryon production and persists in the jet region when the \pttrig is not very high. Studying the strange hadron productions via the two particle correlation method in the future experimental analysis will be of great interest to understand the appearance of strange baryon anomaly behavior observed in small systems and discriminate different physics mechanisms contributing to the strange baryon enhancement at intermediate \pt.

\begin{acknowledgements}

This work is supported by the National Natural Science Foundation of China (12275103, 11875143, 11905188 and 12061141008) and the Innovation Fund of Key Laboratory of Quark and Lepton Physics LPL2020P01 (LZ).

\end{acknowledgements}
%

\bibliographystyle{spphys}       
\bibliography{StrInJetUE}   

\end{sloppypar}
\end{document}